\documentstyle [12pt] {article}

\begin{document}
\title {\large \bf  Representations and Q-Boson Realization of The Algebra
 of Functions on
The Quantum Group  $ GL_{q}(n) $}
\author {V. Karimipour}
\date { }
\maketitle
\begin {center}
Institute for studies in Theoretical Physics and Mathematics {\large \bf
(IPM) }
\\ P.O.Box 19395-5746 Tehran, Iran \\
{\it  Department of Physics , Sharif University of Technology\\
P.O.Box 11365-9161 Tehran, Iran\\
Email. Vahidka@irearn.bitnet}.
\end{center}
\vspace {10 mm}
\begin {abstract}
We present a detailed study of the representations of the algebra of
 functions
on the quantum group $ GL_q(n) $. A q-analouge of the root system is
constructed
for this algebra which is then used to determine explicit matrix
representations of the generators of this algebra.
At the end a q-boson realization
of the generators of $ GL_q(n) $ is given.
\end{abstract} \newpage
\noindent
{\large \bf 1.  Introduction}\\

Although a lot of results exist concerning representations of the quantum
algebras [ 1-5 ], besides some general theorems [6-7], very few explicit
representations have been constructed [8-12] for the dual objects, thats
the
quantum
matrix algebras or more presizely the deformation of the algebra of
 functions
on the group.
In a recent letter [13] we outlined the general method for construction
of the
finite dimensional representations of the quantum matrix group $ GL_q(n) $
( more presizely the quantization of the algebra of functions on $ GL_q(n)$
).
It was proved in [13] , that
finite dimensional irreducible  representations  of this
algebra exist only when $ q $ is a root of unity ( $ q^p = 1 $ ) and the
dimensions of these representations  can only be
one
of the
following values: $ {p^N \over 2^k } $ where $ N = {n(n-1)\over 2 } $
and $ k
\in \{ 0, 1, 2, . . . N \} $.The topology of the space of states was also
clarified ( see also prop. 8 of the present article ).
The method developed in [13]  was based on the introduction of a certain
subalgebra of $ GL_q(n) $ denoted by $ {\bf \Sigma}_n $ for which one could
construct finite dimensional representations in a very straightforward way.
This subalgebra is in fact nothing but a nice root decomposiotion of the
original algebra. It was then shown that from each irreducible $
{\bf \Sigma}_n
$ module one can construct an irreducible $ GL_q(n) $ module.
This strategy has already been  carried out by the present author for the
quatum
groups $ GL_{q,p}(2) $ [9] and $ GL_q(3) $ [10-11] .
What has remained to be done however is the explicit construction of the
general $
{\bf \Sigma}_n $ modules in all its details. This is the subject of the
present
letter.\\
Two basic steps in this construction are:\\
i) A further redefinition of the generators of $ {\bf \Sigma}_n $ such
that all
the roots decouple into mutually commuting pairs ( see eq. (24) ). \\
ii) Introduction of a new identity concerning the quantum determinants(eq.
(40 ) ) which paves the way for the determination of the weights of
representations.\\

{\large \bf 2. The Root system of $ GL_q(n) $}\\

The quantum matrix algebra $ GL_q(n) $ [14]
is a Hopf algebra generated by unity
and the elements $ t_{ij} $ of an $ n\times n $ matrix T, subject to the
relations [15]:\\ \begin{equation} R \ T_1T_2 = T_2 T_1 R \end {equation}
where R is the solution of the Yang-Baxter equation corresponding to $
SL_q(n) $ [16]:
$$ R = \Sigma_{i\ne j} e_{ii}\otimes e_{jj} + \Sigma_i q e_{ii}\otimes
e_{ii}
+ ( q-q^{-1} ) \Sigma_{i<j}e_{ji}\otimes e_{ij} $$
The commutation relations derived from (1) can be neatly expressed in
the following way.\\
For any for elements $ a, b, c, $ and $ d $ in the respective
positions specified by rows and columns (ij) ,(ik),(lj) and (lk),
the following relations hold :
$$ ab =q ba \hskip 1cm cd = q dc  $$
$$ ac =q ca \hskip 1cm bd = q db  $$
$$ bc=cb \hskip 1cm ad-da=(q - q^{-1})bc $$
For any matrix $ T \in GL_q(n) $ , a quantum determinant $ D_q ( T ) $ is
defined with the properties:
$$ [ D_q T, t_{ij} ] = 0  \hskip 2cm \forall t_{ij} \in T $$
$$ \Delta D_q ( T ) =  D_q ( T ) \otimes D_q ( T ) $$
The quantum determinant of T acquires a natural meaning as the q-analogue
of the  volume form when the quantum group is considered as the
automorphism
group on the quantum vector space associated to $ GL_q(n) $ [17] .
It has the following explicit expression: \begin{equation} D_q(T) =
\Sigma_{i=1}^{n}(-q)^{i-1}t_{1i}\Delta_{1i}\end{equation}
where $\Delta_{{1i}} $ is the q-minor corresponding to $ t_{1i}$  and is
defined by a similar formula.

In eq. 2 $ D_q(T) $ has been expanded in terms of the elements in the
first
row of T . Another useful expansion is in term of the last column of T:
\begin{equation} D_q(T) =
\Sigma_{i=1}^{n}(-q)^{n-i}\Delta_{in} t_{in}\end{equation}

To proceed toward constructing the root system of $ GL_q(n) $
let us label the elements
of
the matrix T as follows:\\
$$ T = \left( \begin{array}{llllllll} .&.&.&.&.& .& Y_1 & H_1
\\.&.&.&.&.&  Y_2 & H_2 & X_1 \\.&.&.&.& Y_3 & H_3 & X_2 & . \\.&.&.&Y_4&
 H_4
&X_3
&.&.\\.&.&.&.&.& .&.&.\\.&.&.&.&.& .&.&.\\Y_{n-1}&H_{n-1}& X_{n-2}
&.&.&.&.&.\\
H_n& X_{n-1} .&.&.&.&.& .
\end{array} \right)$$\\
Consider the elements $H_i ,X_i $ and $ Y_i $ together with the q-minors
(q-determinants of the submatrices)\\
$$ H_{ij} = det_q \left( \begin{array}{llll} .&. &.& H_i \\
.&.&.&. \\
.&.&.&.\\ H_j &.&.&.\end{array} \right)$$
$$ X_{ij} = det_q \left( \begin{array}{llll} .&. &.& X_i \\
.&.&.&. \\ .&.&.&.\\ X_j &.&.&.\end{array} \right)$$
$$ Y_{ij} = det_q \left( \begin{array}{llll} .&. &.& Y_i \\
.&.&.&. \\ .&.&.&.\\ Y_j &.&.&.\end{array} \right)$$
For convenience we sometimes denote $H_i,X_i $ and $ Y_i $ by $ H_{ii},
X_{ii}
$ and $ Y_{ii} $ respectively.\\
The subalgebra $ {\bf \Sigma _n } $ \ is equal to\  $ {\bf \Sigma _n^0 }
\oplus
{\bf \Sigma _n^+ } \oplus {\bf \Sigma _n^- } $ \ \ where the latter are
 generated respectively by the
elements $ H_{ij}  i\leq j , X_{ij}   i\leq j $ and $ Y_{ij} i\leq j
$\\

We call the elements $ X_i $ and $ Y_i
$ simple roots and the elements $  X_{ij}\ \  \ \  i < j  $ and $ Y_{ij}
,\ \ \  i < j
$ non-simple roots. As will be shown below  the generators $ H_{i} $
will
play the role of Cartan subalgebra elements and
the elements $ X_{ij}\ \ \  i\leq j  $ ( resp.$ Y_{ij} \ \ \ i\leq j  $)
will act as raising and
lowering operators.
We use the word root in a special sence, by which we mean that from
representations of roots, representations of all the other elements of the
quantum group can be constructed.
For $ GL_q(n) $ there are $ N = { n(n-1)\over 2 }$ pair of positive and
negative roots.

The reason why constructing $ {\bf \Sigma}_n$ modules is easy is due to the
very crutial
fact that almost all the relations  between generators of ${\bf
\Sigma }_n$ are
multiplicative or of Heisenberg-Weyl type.
By multiplicative relation between two element $ {\bf x} $ and $ {\bf y} $,
we
mean a relation of the form $ {\bf x y  } = q^{\alpha }  {\bf y x } $ ,
where $ {\alpha }$ is an integer.\\
{\bf Remark}: In the rest of this paper a multiplicative relation between
${\bf x } $ and $ {\bf y } $  is indicated as $ {\bf x y } \approx
{\bf y x }
$ \\
The important properties of
$ {\bf \Sigma }_n$ is encoded in the following propositions ( see ref. [10]
for their proof):\\
{\bf proposition 1:} \ \
For all $ i, j, k, $\ \ and \ \ $ l $ :
\begin{equation}[H_{ij},H_{kl}] = 0\end {equation}
\begin{equation}[X_{ij},X_{kl}] = 0\end {equation}
\begin{equation}[Y_{ij},Y_{kl}] = 0\end {equation}  \\

Thus ${{\bf \Sigma }_n}^{0} $ \ \ and  \ \ $  {{\bf \Sigma }_n}^{\pm} $
are three commuting subalgebras of $ GL_q(n) $ .
For the relations between the generators of $ {\bf\Sigma^0 } $ and $ { \bf
\Sigma^{\pm} } $  we have:\\
{\bf proposition 2}\\
\begin{equation} H_i X_{ij} = q X_{ij}H_i \hskip 1cm
\forall j \geq i \end{equation}
\begin{equation} H_{j+1} X_{ij} = q X_{ij}H_{j+1} \hskip 1cm \forall
 i\leq j  \end {equation}
\begin{equation} H_k X_{ij} =  X_{ij}H_k \hskip 1cm  k \ne i , j+1\end
{equation}
\begin{equation} H_{ij} X_{kl} \approx  X_{kl}H_{ij}
\hskip 1cm \forall i,j,k,l\end {equation}
with $ (q\longrightarrow q^{-1}  , X_{ij}\longrightarrow Y_{ij} ) $\\

{ \bf Remark:} The exact coefficients in
relation ( 10 )  can easily be determined ( see  Lemma 10 of
ref. [10] ). We need in particular the relations:
\begin{equation} \ \ \ \ \ \ H_{ij} X_{k} =  X_{k}H_{ij} \hskip 2cm i\leq
 k \leq j-1 \end {equation}
\begin{equation} H_{ij} X_{ij} = q X_{ij}H_{ij} \ \ \ \end {equation}
\begin{equation} H_{i+1,j+1} X_{ij} = q X_{ij}H_{i+1,j+1}  \end {equation}
\begin{equation} [ H_{i,j+1}, X_{ij} ] =  [ H_{i+1,j}, X_{ij} ] = 0
\end {equation} \\

{\bf The relations
between elements of $ {{\bf \Sigma}^+}_n $ and $ {{\bf \Sigma}^-}_n $}.\\

{\bf proposition 3: }\\
\begin{equation} Y_{kl} X_{ij} \approx X_{ij}Y_{kl} \hskip 1cm
(k,l)\ne(i,j)\end {equation}
\begin{equation} Y_{i} X_{i} - X_{i}Y_{i} = (q - q^{-1} ) H_{i}H_{i+1}
\end {equation}
\begin{equation} q^{-1} Y_{ij} X_{ij} - q X_{ij}Y_{ij} = (q^{-1}-q)
H_{i,j+1}H_{i+1,
j}\end {equation} \\

{\bf proposition 4}.\ \  For $ q^p=1$ the p-th power of all the
elements of ${\bf \Sigma}_n$ are central.\\

{\bf Proof :}
For the multiplicative relations this is obvious. The only
non-multiplicative relations
are (16) and (17) . From (16) we have: \begin{equation} H_i H_{i+1} X_i
= q^2 X_i H_i
H_{i+1} \end {equation}  using this relation and (16) we find by induction:
\begin{equation} Y_i X_i^n = X_i^n Y_i + ( q-q^{-1} ) \big\lbrace
{ q^{2n} - 1 \over q^2 - 1
} \big\rbrace X_i^{n-1} H_i H_{i+1} \end {equation}  which shows that for
$ q^p = 1 $
\begin{equation} Y_i X_i^p = X_i ^P Y_i \end {equation}
A similar argument shows that $ Y_i X_i^p = X_i ^p Y_i $

For the relation (17) we use the fact that $ H_{i,j+1}H_{i+1,j} X_{ij} =
X_{ij} H_{i,j+1}H_{i+1,j} $\ \ .By induction from (17) we obtain:
\begin{equation}  Y_{ij} X_{ij}^n = q^{2n} X_{ij}^n Y_{ij} + ( 1 - q^{2n}
 ) X_{ij}^{n-1} H_{i,j+1} H_{i+1,j} \end {equation}
which again shows that :
\begin{equation} [ Y_{ij}, {X_{ij}}^p ] =  [ {Y_{ij}}^p , {X_{ij}} ] = 0
\end {equation}
\\

Let ${\bf V}$ be a $ {\bf\Sigma}_n$ module. We call this module trivial
if, the
 action of
one or more of
the elements of ${\bf \Sigma}_n$ on it, is identically zero. We are
interested in
nontrivial
$ {\bf\Sigma}_n$ -modules. (the trivial one's are representations of
reductions
of ${\bf \Sigma}_n$ ).\\

{\bf proposition 5}.\ \ A $ {\bf\Sigma}_n$ module $ {\bf V} $ is
nontrivial only if
all the subspaces
$$ {\bf K}_{ij}\equiv \{ \vert v > \in V \vert  \hskip 1cm H_{ij}
\vert v  > =
0 \}$$ are zero dimensional.\\ \\
{\bf Proof}  :  Suppose that for some $ i $ and $ j $ \ \ $ dim \ \
{\bf K}_{ij}
\ne 0 $. We choose a basis like
$ \{ \vert {\bf e}_{i} >  , i = 1,...N \} $ for $ {\bf K}_{ij} $.
Due to the multiplicative relation of $ H_{ij} $ with all the elements of $
{\bf \Sigma}_n $ it is clear that for any $ {\bf m } \in {\bf \Sigma }_n$
we have:
$$ H_{ij} \ \  {\bf m }\ \ \vert {\bf e}_k > \ \ \approx  {\bf m }\ \
H_{ij}
\ \ \vert {\bf e}_k >  \ = 0 $$
Therefore $ {\bf m }{\bf e}_k \in K_{ij} $ which means that the basis
vectors
$ e_k $ transform among themselves under the action of $ {\bf \Sigma}_n $ .
Since V is assumed to be irreducible we have $ K_{ij} = V $
and $$ H_{ij}\  V = H_{ij}\  K_{ij} = 0 $$
which shows that V is a trivial ${\bf \Sigma}_n $ module.\\
\\
{\bf proposition 6}:\\  {i -}  Finite dimensional irreducible
representations
 of $ {\bf
\Sigma }_n $  exist only when q is a root of unity.\\ {ii-} Any non-trivial
${\bf \Sigma }_n$ module {\bf V} is also an $ {\bf GL_q(n)} $  module and
vice
versa.\\ \\
{\bf proof}: The proof of this proposition is exactly parallel to the
case of
$ GL_q(2)$ [9] and  $ GL_q(3) [10]$. One uses the expressions (2)(resp. 3)
 for
the q-determinants
$ Y_{ij} $(resp. $ X_{ij}$ (starting
from $ j=i+1 $,continuing to $ j=i+2,i+3 ... $) and uses the fact that
in the representation of $ {\bf \Sigma}_n$, all the
elements $ H_{ij} $ are invertible diagonal matrices.
As an example,in the appendix we carry out this procedure explicitly for
the
quantum group $ GL_q ( 4 ) $.
Note that
invertibility of $H_{ij}$'s
(due to proposition. 5) is crutial here, otherwise one can not define the
actions of the remaining elements of T or V.\\ \\
{\large \bf 3. Representations}\\

To develop the full representation theory we rescale the roots as follows:
\begin{equation} {\bf h}_{ij} = H_{ij} \hskip 2cm {\bf x}_{ij} = {\mu_{ij}}
^{{-1\over 2}}
X_{ij} \hskip 2cm {\bf y}_{ij} = {\mu_{ij}}^{{-1\over 2 }} Y_{ij}
\end {equation} \\

where $ \mu_{ij} = ( H_{ij}H_{i+1,j+1}) $\\

As the reader can verify, with this redefinition the root system is
completely disentangled into
msutually
commuting pairs, while all the relations between $ H_{ij}$ and $ X_{ij} $
( $ Y_{ij} $ ) remain intact.
Instead of (15-17) one will have:

\begin{equation} \ \ \ [ {\bf x}_{ij} ,{\bf y}_{kl} ] = 0 \hskip 2cm (k,l)
\ne (i,j) \end {equation}
\begin{equation} q^{-1} {\bf x}_i {\bf y}_i - q {\bf y}_i {\bf x}_i =
q^{-1} - q \end {equation}
\begin{equation} [ {\bf x}_{ij} , {\bf y}_{ij} ] = ( q-q^{-1} )
{ {\bf h}_{i,j+1}
{\bf h}_{i+1,j}\over {\bf h}_{ij} {\bf h}_{i+1,j+1} } \end {equation}
{}From these relations one can also obtain the more general relations :
\begin{equation} {\bf y}_i {{\bf x}_i}^l = q^{-2l} {{\bf x}_i}^l {\bf y}_i
 + ( 1 - q^{-2l} )
{{\bf x}_i}^{l-1}
\end {equation}  \begin{equation} {\bf y}_{ij} {{\bf x}_{ij}}^l =
{{\bf x}_{ij}}^l{\bf y}_{ij} + q (
q^{-2l} - 1 ) {{\bf x}_{ij}}^{l-1} { {\bf h}_{i,j+1} {\bf h}_{i+1,j}\over
{\bf h}_{ij} {\bf h}_{i+1,j+1} } \end {equation} \\

With this redefinition the only structure constants of the algebra are
the coefficients
between the $ {\bf h}_{ij} $ and $ {\bf x}_{ij} $ .Table 1 shows these
 structure constants for
$ GL_q (4) $.\\

Consider a common eigenvector of $ {\bf h}_{ij} $'s which we denote by
$ {\bf \vert 0 >} $ with eigenvalues $ {\bf h}_{ij} {\bf \vert 0 >} =
\lambda_{ij} {\bf \vert 0 >}$
and construct an $ N = { n(n-1)\over 2}  $ dimensional hypercube of states
\begin{equation} W = \{ \ \ \vert {\bf l } >   = \prod_{i,j} ({\bf x}_{ij}
)^{l_{ij}}
\vert {\bf 0 }>\hskip 1cm  0 \leq l_{ij}
\leq p-1\ \  \} \end {equation} \\
where $ {\bf l } $ is a vector $ {\bf l } = \sum_{i \leq j} l_{ij} {\bf
e}_{ij} $ in the lattice.
{}From (10) all the states of $ W $ are eigenstates of $ {\bf h}_{ij} $ 's.
\begin{equation} {\bf h}_{ij}\vert {\bf l }> = q^{c_{ij}({\bf l })}
 \lambda_{ij} \vert {\bf l }>
\end {equation}
The parameters $ c_{ij}( {\bf l } ) $ can be  easily calculated by using
 the
sructure constants.( see the  appendix where the case of $GL_q(4)$ is
considered as an example)\\

Each positive root generates one direction of this hypercube.Because
 of (5) we have:
\begin{equation} {\bf x}_{i} \vert {\bf l }> = \vert {\bf l} + {\bf e}_{i}>
\end{equation}

\begin{equation} {\bf x}_{ij} \vert {\bf l }> = \vert {\bf l} +
{\bf e}_{ij} >
\end{equation} Since $ {{\bf x}_{ij}}^p  $  is central we can set it's
value on
$ W $ equal to a c-number $ \eta_{ij} $ . Therefore we have :
\begin{equation} {\bf x}_{ij} \vert {(p-1){\bf e}_{ij}} > \ \
=\ \ \eta_{ij} \vert
{\bf  0}> \end{equation} The last relation says much more.
We need some terminology. Denote by $ F_{ij}^0 $ and $ F_{ij}^1 $
the two faces
which are perpendicular to the vector $ {\bf e}_{ij} $ respectively passing
through the origin and the point $ (p-1) {\bf e}_{ij} $ .
Now if $ {\bf v } $ is any vector in $ F_{ij}^1 $ then by eq. (10) we have:
\begin{equation} {\bf x}_{ij} \vert (p-1){\bf e}_{ij} +{\bf v} > =
\eta_{ij}
\vert {\bf v }> \end{equation}
In this way when $ \eta_{ij} $ is nonzero the generator $ {\bf x}_{ij} $
folds
each face $ F_{ij}^1 $ onto the face $ F_{ij}^0 $ .
Define the action of $ {\bf y}_{ij} $ on $ \vert {\bf 0 } > $ by:
\begin{equation} {\bf y}_{ij} \vert {\bf 0 }> =\alpha_{ij} \vert (p-1){\bf
e}_{ij}> \end{equation}
By the same reasoning as in the case of $ {\bf x}_{ij} $ one can show that
when $ \alpha_{ij} $ is nonzero the generator $ {\bf y}_{ij} $ folds
the face $
F_{ij}^0 $ onto
the face $ F_{ij} ^1 $, i.e: for any vector $ {\bf u } $ lying in
 $ F_{ij} ^0 $
\begin{equation} {\bf y}_{ij} \vert {\bf u }> =\alpha_{ij} \vert (p-1){\bf
 e}_{ij} +{\bf u }> \end{equation}\\
We now  calculate the action of the negative roots on the other
states of $ W $ .
Thanks to the commutation relations (24) one can calculate the action
of any root like $ y_k $ on any state as follows:
\begin{equation} {\bf y}_k \vert {\bf l } > = {\bf y}_k ( \prod_i
{\bf x}_i^ {  \ l_i} )
\vert
{\bf 0 } > = \prod_{i\ne k } {\bf x}_i^ { \ l_i}\  y_k \  {{\bf x}_k}^ {
\ l_k} \vert {\bf 0 }> \end {equation}
For simplicity of notation, in this equation we have represented any
 positive
( resp. negative ) root by the symbol $ {\bf x}_k $ ( resp. $
{\bf y}_k $ ) and have not
distinguished between simple and nonsimple roots.
One then uses eqs. (25-26) to complete the calculation. The result is:

\begin{equation} {\bf y}_{i} \vert {\bf l } > = ( q^{-2l_i} \alpha_i
 \eta_i + (
1-q^{-2l_i}))\vert {\bf l - {\bf e }}_{i}  > \end {equation}
\begin{equation} {\bf y}_{ij} \vert {\bf l } > = ( \alpha_{ij}
\eta_{ij} + q (
1-q^{-2l_{ij}}) s_{ij})\vert {\bf l - {\bf e }}_{ij}  > \end {equation}
where $ s_{ij} = { \lambda_{i,j+1} \lambda_{i+1,j} \over \lambda_{ij}
\lambda_{i+1,j+1}} $
This shows that each $ {\bf y}_{ij} $ acts as a lowering operator in the
direction $ {\bf e}_{ij} $ of the hypercube.\\ \\

It remains to determine the parameters $ \lambda_{ij} $.
Clearly calculation of these parameters by direct expansion of
$ {\bf h}_{ij} $
is cumbersome. Instead we proceed as follows: Denote by $ E_{i,j} $ (resp.
$ E_{ij,kl} $)  the q-minors obtained from a quantum matrix $ E $ by
deleting\  the \ rows \ $ i $ \ ( resp.\  $ i $ \ and \ $j $)\  and \ columns\
k ( resp. $ k $ and
$ l $ ). \ Then we conjecture that the following identity is true:
\begin{equation} E_{jl}E_{ik} - qE_{jk}E_{il} = E_{ij,kl}
Det_q E \end {equation}
The classical limit of this identity is well known. In the quantum case it
can be checked by direct computation for low dimensional $ GL_q ( n )  $
matrices. Later on we will give further justification for it using the
conjucation properties of $ {\bf \Sigma}_n $.
We now use this relation to determine the parameters $ \lambda_{ij} $.
Eq. (40
) implies the following relation  in $ {\bf \Sigma}_n $:
\begin{equation} Y_{ij} X_{ij} = q  H_{ij} H_{i+1,j+1} + H_{i,j+1}
H_{i+1,j} \end {equation}

Further justification is obtained by using the conjucation properties of
T as follows.
For $ q $ on the unit circle the elements of T allow the following
conjucation:
\begin{equation} t_{ij}^{\dagger} = t_{ij} \end {equation}
This results in the following conjucation properties in $ {\bf \Sigma}_n$:
\begin{equation} X_{ij}^{\dagger} = X_{ij}  \hskip 2cm Y_{ij}^{\dagger} =
Y_{ij} \hskip
 2cm H_{ij}^{\dagger} = H_{ij} \end {equation}
One can then conjucate both sides of this equation to obtain:
\begin{equation} X_{ij} Y_{ij} = q ^{-1} H_{ij} H_{i+1,j+1} +
 H_{i,j+1} H_{i+1,j} \end {equation}
Combination of eqs. (41) and (44) then leads to eq. (17) which has
 already been
proved in [10] .

In terms of the rescaled generators relation (41) takes the following
form  :

\begin{equation} {\bf x} _i {\bf y} _i =  1 + q  {{\bf h}_i {\bf h}_{i+1}\over
{\bf h}_{i,i+1}} \end {equation}
\begin{equation} {\bf x}_{ij}{\bf y}_{ij} = 1 + q  { {\bf h}_{i,j+1}
{\bf h} _{i+1,j}\over {\bf h}_{i,j}
 {\bf h}_{i+1,j+1} }\end {equation}
Now these relations help us to determine the parameters $ \lambda_{ij} $:
Acting on the state $ {\bf \vert 0 }> $ by both sides of (45,46)
we obtain :
\begin{equation} \alpha_i \eta_i = 1 + q { \lambda_{i,i+1}\over \lambda_i
 \lambda_{i+1} }\end {equation}
\begin{equation} \alpha_{ij} \eta_{ij} = 1 + q  { \lambda_{i,j+1}
 \lambda_{i+1,j} \over
\lambda_{ij} \lambda_{i+1,j+1}} \end {equation}          \\
or
\begin{equation} \lambda_{i,i+1} = q^{-1} \lambda_i \lambda_{i+1}
( \alpha_i \eta_i - 1 ) \end {equation}
\begin{equation} \lambda_{i,j+1} = q^{-1} { \lambda_{i,j}
\lambda_{i+1,j+1}(  \alpha_{ij}
\eta_{ij} - 1 ) \over \lambda_{i+1,j}}\end {equation} \\

Let us call $ \lambda_{ij} $ the weights of the representation and
call each
$ \lambda_{i,i+k} $ a weight
at level k. Eqs. (49-50) express the weights at each level in terms of
the weights at
the lower level.(see the appendix for the example of $ GL_q(4) $ )\\ \\

{\large \bf 4. Types of Representations}\\

We complete our analysis of
representation of $ GL_q(n)
$ by a discussion on the various types of representations.
Each representation is defined by the  $ n^2 $ parameters\  $ \alpha_{ij},
\ \ \eta_{ij} \ \ $ and\  $ \lambda_{i} $. The type of representation
depends on
the values of the parameters $ \alpha_{ij}$ and $ \eta_{ij} \ \ \ $.More
presizely we have:\\ \\
{\bf Proposition 8 :}
The dimensions of the irreducible  representations  of $ GL_q(n) $ can
only be
one
of the
following values: $ {p^N \over 2^k } $ where $ N = {n(n-1)\over 2 } $
and $ k
\in \{ 0, 1, 2, . . . N \} $
For each $ k $ the topology of the space of states is $ (S^1)^{\times(N-k)}
\times [ 0 , 1 ] ^{\times (k)} $ (i.e. an $ N $ dimensional torus for
$ k=0 $
and an $ N $ dimensional cube for $ k = N $ ).\\

{\bf Proof: }  Our style of proof is a generalization of the one given in
[8] and [10] for the case of $ GL_{q,p}(2) $ and $ GL_{q}(3) $
respectively.\\
Let {\bf V} be an $ GL_q ( n ) $ module with dimesion d .
Depending on the values of the parameters $ \alpha_{ij} $  and
$ {\eta_{ij} }$
three cases can happen:\\ \\
Case {\bf a}: \ \ \ $ \alpha_{ij} \ne 0 \ne \eta_{ij} \hskip 2cm
\forall i,j $\\

In this case $ d $ can not be greater than $ p^N $, otherwise the cube
$ W $ will span an invariant submodule
which contradicts the irreducibility of V . The dimension of V can
not be less than $ p^N $  either
since this means that the lenght of one of the sides of the cube W
(say in the $ i$-th direction )
must be less than p . Therefore there must exist a positive integer $ r<p
$ such that $
{{\bf x}_i}^r \vert 0 > = 0 $ which means that $ \eta_i \vert 0 > =
{{\bf x}_i}^{p-r}{{\bf x}_i}^r
\vert 0 > = 0 $
contradicting the original assumption.
The topology of the space of states in this case is an  $N$ dimesional
torus
( ${ S^1}^{\times N } $ )\\

Case {\bf b}:\ \  For some $ ( ij ) $ $ \alpha_{ij} \ne 0 , $ but
$ \eta_{ij}= 0  $
or vice versa:\\

 In this case the representation is semicyclic in the $ {ij}
 $ direction.\\

Case {\bf c}:\ \ for some ( ij ) $ \alpha_{ij} = \eta_{ij} = 0  $:\\ \\
In this case the representation has a highest and a lowest wieght in
the $ {ij}$-th direction.\\

If\ \  $ d <  P^N $\   there must exist an integer like $ r<p$ such that
$ {\bf x}_{ij}^r \vert 0 > = 0 $ and $ {\bf x}_{ij}^l\vert 0> \ne 0 $ for
$ l\ < \ r  $. Now denote $ {\bf x}_{ij}^r\vert 0> $ by $ {\bf u}_0 $ and
 consider the
string of states $ {\bf y}_{ij}^{l'}\ {\bf u}_0 $ .This string of states
 must terminate somewhere. Thats there must exists
 an integer
like $ r'$ such that $ {\bf y}_{ij}^{r'}\ {\bf u}_0 = 0 $ and $ {\bf y}_
{ij}^{r'-1} {\bf u}_0 \ne 0 $
Therefore $$ 0 = {\bf x}_{ij}{\bf y}_{ij}^{r'}u_0 = ( {\bf y}_{ij}^{r'}
{\bf x}_{ij} + q (
q^{-2r'}-1){\bf y}_{ij}^{r'-1}
{ {\bf h}_{i,j+1} {\bf h}_{i+1,j}\over
{\bf h}_{ij} {\bf h}_{i+1,j+1} })
{\bf u}_0 =
q(q^{-2r'}-1)s_{ij}\ {\bf u}_0 $$
which means that $ q^{2r'}=1 $ or $ r'={p\over 2 } $ . r' is in fact the
lenght
of the edge of the cube W in the $ ij-th $ direction , the other
dedges being
of
lenght p . The dimension of V is in this case $ { p^N\over 2 } $ . The
topology
of the space of states is in this case
$ [0 , 1]\times {S^1}^{\otimes N-1} $. By repeating this analysis for other
pairs of the parameters  the assertion is proved.\\ \\

{\large \bf Q-Boson Realization}\\ \\
One \ can \ construct \ an \ infinite \ dimensional \ representation
\ ( q - analouge\  of Verma
Module\  ) by setting all $ \alpha_i = 0  $ and relaxing all the conditions
 of preiodicity  . It is
then very easy to determine the q-boson realization of all the
generators of
${\bf \Sigma}_n $ and hence of $ GL_q(n) $.

The q-boson algebra [18-20] $ B_q$ is generated by three elements $ a ,
a^{\dagger}$ , and $ N $ satisfying the relations:
\begin {equation} a a^{\dagger} - q^{\pm 1 } a^{\dagger} a  = q^ {\mp N }
\end{equation}
\begin {equation} q^{\pm N }a = q^{\mp1} a q^{\pm N} \hskip 2cm   q^{\pm N}
a^{\dagger} = q^{\pm 1} a^{\dagger} q^{ \pm N } \end{equation}
A more useful form of the algebra is obtained if one replaces
the above equations by
the following pair of relations:
\begin{equation} a a^{\dagger} = [ N + 1 ] \hskip 2cm     a^{\dagger}
 a  = [ N ] \end {equation}
where the symbol $ [ N ] $ as usual stands for ${ q^N - q^{-N}
\over q-q^{-1}}$
for N being a number or an operator.\\

On the q-Fock space $ F_q$ spanned by the states $ \vert n > \equiv
{a^{\dagger}}^n \vert 0 > $ the action of the generators are :
\begin{equation} a^{\dagger} \vert n > =  \vert n+1 >  \end {equation}
\begin{equation} a  \vert n > = [n]_q \vert n-1 > \end {equation}
\begin{equation} N \vert n > = n \vert n >   \end {equation}

Consider $ N $ commuting q-bosons ( i.e. $ a_i ,
{a^{\dagger}}_i , N_i ; i =
1... N $ ) and their representation on the q-Fock space
$ F_q^{\otimes N }$.
Then
if $ \Psi $ is the natural isomorphism from W to  $ F_q^{\otimes N }$,
satisfying:
\begin{equation} \Psi : \vert {\bf l} > \longrightarrow
\prod_{i=1}^{N} {a_i}^{l_i}
\vert 0 >\end {equation}
the induced representation $ \Psi $ is defined by [13] :
\begin{equation} \Psi^* (g ) = \Psi \circ g \circ \Psi^{-1}  \hskip 2
cm \forall g \in End \
\ W\end{equation}
We will then have the following $ n^2$ parameter family of
q-boson realization
of the quantum group $ GL_q(n) $.\\ \begin{equation} {\bf x}_i = a_i
^{\dagger} \hskip 2cm {\bf x}_{ij} =
a_{ij} ^{\dagger} \end {equation}

\begin{equation} {\bf y}_i = ( q-q^{-1} ) a_i q^{-N_i} \hskip 2cm
{\bf y} _{ij} = q ( q^{-1} - q ) s_{ij}a_{ij} q^{-N_{ij}} \end {equation}
\begin{equation} {\bf h}_i = \lambda_i q^{C_i(N)} \hskip 2cm h_{ij} =
q^{C_{ij}(N)}\lambda_{ij} \end {equation}
\\ \\
{\large \bf Acknowledgement :}
I would like to thank all my colleagues in the physics department of IPM for
very valuable discussions.I also espress my sincere thanks to A. Morosov
for very interesting comments made during his visit to IPM.\\ \\

{\large \bf Appendix } An Example : The Case of $ GL_q(4) $\\

The structure constants of $ GL_q(4) $ ( see [ 13 ] ) are indicated
in table 1. Conseqently we obtain  the following actions:

$$ h_{1} \vert {\bf l} > = q^{l_1 + l_{12} + l_{13} }\lambda _{1}
\vert {\bf l } > $$
$$ h_{2} \vert {\bf l} > = q^{l_1 + l_{2} + l_{13} }
\lambda _{2} \vert {\bf l } > $$
$$ h_{3} \vert {\bf l} > = q^{l_2 + l_{3} + l_{12} }\lambda _{3}
\vert {\bf l } > $$
$$ h_{4} \vert {\bf l} > = q^{l_3 + l_{13} + l_{23}  }\lambda _{4}
\vert {\bf l } > $$
$$ h_{12} \vert {\bf l} > = q^{l_2 + l_{12} + l_{23} + l_{13} }
\lambda _{12}
\vert {\bf l } > $$
$$ h_{23} \vert {\bf l} > = q^{l_1 + l_{3} + l_{12} + l_{23} }\lambda _{23}
\vert {\bf l } > $$
$$ h_{34} \vert {\bf l} > = q^{l_2 + l_{12} + l_{23} + l_{13} }
\lambda _{34}
\vert {\bf l } > $$
$$ h_{13} \vert {\bf l} > = q^{l_3 + l_{23} + l_{13} }\lambda _{13}
\vert {\bf l } > $$
$$ h_{24} \vert {\bf l} > = q^{l_1 + l_{12} + l_{13} }\lambda _{24}
\vert {\bf l } > $$

The weights $ \lambda_{ij} $ are determined from (49-50) to be:

$$ \lambda _{12} = q^{-1}\lambda_1 \lambda_2 ( \alpha_1 \eta_1 - 1 ) $$
$$ \lambda_{23} = q^{-1}\lambda_2 \lambda_3 ( \alpha_2 \eta_2 - 1 ) $$
$$ \lambda_{34} = q^{-1}\lambda_3 \lambda_4 ( \alpha_3 \eta_3 - 1 ) $$
$$ \lambda _{13} = q^{-1}{\lambda_{12} \lambda_{23} ( \alpha_{12}
\eta_{12} - 1 )\over \lambda_2 } $$
$$ \lambda _{24} = q^{-1}{\lambda_{23} \lambda_{34} ( \alpha_{23}
\eta_{23} - 1 )\over \lambda_3 } $$
$$ \lambda _{14} = q^{-1}{\lambda_{13} \lambda_{24} ( \alpha_{13}
\eta_{13} - 1 )\over \lambda_{23} } $$  \\

In the following we carry out explicitly the process of reconstruction of
$ GL_q(4) $ from $ {\bf \Sigma}_4 $\\

Let us label the elements of $ T\in GL_q(4) $ as follows:

$$ T = \left( \begin{array}{llll} p&l_1&Y_1&H_1
\\l_2&Y_2&H_2&X_1 \\Y_3&H_3&X_2&m_1
\\H_4&X_3&m_2&n \end{array} \right)$$
Here we have:
$$ X_{12} = H_2 m_1 - q X_1 X_2 $$
$$ X_{23} = H_3 m_2 - q X_2 X_3 $$
$$ Y_{12} = l_1 H_2 - q Y_1 Y_2 $$
$$ Y_{23} = l_2 H_3 - q Y_2 Y_3 $$
{}From which we obtain:
$$ m_1 = {H_2}^{-1} ( X_{12} + q X_1 X_2 ) $$
$$ m_2 = {H_3}^{-2} ( X_{23} + q X_2 X_3 ) $$
$$ l_1 =  ( Y_{12} + q Y_1 Y_2 ){H_2}^{-1} $$
$$ l_2 =  ( Y_{23} + q Y_2 Y_3 ){H_3}^{-1}  $$
We also have:
$$ X_{13} = H_{23} n - q ( Y_2 m_2 - q H_2 X_3 ) m_1 + q^2 X_{23} X_1  $$
$$ Y_{13} = p H_{23} - q  l_1 ( l_2 X_2 - q H_2 Y_3 ) + q^2Y_1 Y_{23}  $$
{}From which we obtain:
$$ n = H_{23}^{-1}\big \lbrace X_{13} + q ( Y_2 m_2 - q H_2 X_3 ) m_1 + q^2
X_{23} X_1 \big \rbrace $$
$$ p = \big\lbrace Y_{13} +  q  l_1 ( l_2 X_2 - q H_2 Y_3 ) + q^2Y_1
Y_{23}\big\rbrace H_{23}^{-1} $$
These equantions show that once the action of $ {\bf \Sigma }_4
$ is known on $
V
$ the action of $ GL_q(4) $ can be determined uniquely.

\newpage
{\large \bf References}
\begin{enumerate}
\item  G. Lusztig , Adv. Math. 70, 237 (1988); Contemp. Math.
82,59 (1989)
\item  M. Rosso , Commun. Math. Phsy. 117, 581 (1988) ; 124, 307 (1989)
\item  R. P. Roche and D. Arnaudon , Lett. Math. Phsy. 17, 295 (1989)
\item  C. De Concini and V. G. Kac , Preprint (1990)
\item  P. Sun and M. L. Ge , J. Phys. A 24, 3731 (1991)
\item  Ya. S. Soibelman ,Leningrad Math J. {\bf 2 }, 161-178 (1991)
\item  Ya. S. Soibelman and L. Vaksman ; Func. Anal. Appl.
{\bf 22 }(3) 170-181 (1988)
\item  M. L. Ge, X. F. Liu , and C. P. Sun : J. Math. Phys. 38 (7) 1992
\item  V. Karimipour ; Q-Boson realization of the Quantum Matrix
Algebra $ M_q(3) $
J. Phys. A. Math. Gen. L957 - L962 (1993).
\item  V. Karimipour ; Representations of the Coordinate Ring of $ GL_q(3)$
Lett. Math. Phys. $ {\bf(28) }$:207-217 (1993)
\item  V. Karimipour ; Representations of the Quantum Matrix Algebra
 $ M_{q,p}(2) $
J. Phys. A. Math . Gen. , in press
\item J. Floratos   ;
Representations of the Quantum Group $ GL_{q}(2) $  for values of $ q $
on the unit circle
Phys. Lett. B. 233 (3,4) 1989
\item  V. Karimipour ; Representations of the Coordinate Ring of
$ GL_q(n) $
IPM Preprint 93-15 , Tehran (1993)
\item  See L.A. Takhtajan , in M. L. Ge and B. H. Zhao ( eds.)
 Introduction to Quantum
Groups and Integrable Massive Models of Quantum Field Theory,
 World Scientific, (1991)
\item  N. Reshetikhin , L. Takhtajan , and L. Faddeev ; Alg. Anal. 1, 1
78 ( 1989) in Russian
\item  M. Jimbo ; Lett. Math. Phsy. 10, 63 ( 1985) ; 11,247 (1986)
\item  Yu. Manin ; Bonn Preprints MPI/91-47 , MPI/91-60(1991)
\item  L.C. Biedenharn ; J. Phys. A , Math. Gen. {\bf 22} L873 (1989)
\item  A. J. Macfarlane ;  J. Phys. A , Math. Gen. {\bf 22} 4551 (1989)
\item  C. P. Sun and H. C. Fu ; J. Phys. A , Math. Gen. {\bf 22}
L983 (1989)
\newpage

{\bf Table 1 - The structure constants of  $GL_q(4) $ }\\

$$ \left( \begin{array}{lllllllllll} &&x_1 & x_2 & x_3 &
 {\bf x}_{12}& {\bf x}_{23}& {\bf
x}_{13} & &\\
 {\bf h}_1  && q&1&1&q&1&q \\   {\bf h}_2  && q&q&1&1&q&1 \\
 {\bf h}_3  && 1&q&q&q&1&1 \\   {\bf h}_4  && 1&1&q&1&q&q \\

 {\bf h}_{12}& &1&q&1&q&q&q&& \\ {\bf h}_{23}&&q&1&q&q&q&1&&
\\ {\bf h}_{34}&&1&q&1&q&q&q&&
\\{\bf h}_{13}&&1&1&q&1&q&q&&\\{\bf h}_{24}&&q&1&1&q&1&q&&
\end{array} \right)$$\\
$$ i.e:\ \ \  {\bf h}_{12} {\bf x}_{12} = q {\bf x}_{12}{\bf h}_{12} $$

\end{enumerate}
\end{document}